\documentclass{aa}  
\usepackage{natbib}
\usepackage{graphicx}
\usepackage{multirow}
\usepackage[table]{xcolor}
\usepackage{xcolor,colortbl}
\definecolor{lightgray}{gray}{0.9}
\usepackage{txfonts}
\usepackage{aalongtable}
\begin{document}

\title{Dust as interstellar catalyst -- I.}
\subtitle{Quantifying the chemical desorption process}
\author{M. Minissale\inst{1}, F. Dulieu\inst{1}, S. Cazaux\inst{2}, S. Hocuk\inst{3}} 
\offprints{cazaux@astro.rug.nl}
\institute{LERMA, Universit\'e de Cergy Pontoise, Observatoire de Paris, PSL Research University, Sorbonne Universit\'es, UPMC Univ Paris 06, CNRS UMR 8112, 5 mail Gay Lussac, 95000 Cergy Pontoise Cedex, France. 
\and
Kapteyn Astronomical Institute, University of Groningen, P.O. Box 800, 9700AV Groningen, The Netherlands
\and
Max Planck Institute for Extraterrestrial Physics, Giessenbachstr. 1, D-85741, Garching, Germany}

\date{Received ; accepted }

\abstract {The presence of dust in the interstellar medium has profound consequences on the chemical composition of regions where stars are forming. Recent observations show that many species formed onto dust are populating the gas phase, especially in cold environments where UV and CR induced photons do not account for such processes.}
{The aim of this paper is to understand and quantify the process that releases solid species into the gas phase, the so-called chemical desorption process, so that an explicit formula can be derived that can be included into astrochemical models.}
{We present a collection of experimental results of more than 10 reactive systems. For each reaction, different substrates such as oxidized graphite and compact amorphous water ice are used. We derive a formula to reproduce the efficiencies of the chemical desorption process, which considers the equipartition of the energy of newly formed products, followed by classical bounce on the surface. In part II we extend these results to astrophysical conditions.}
{The equipartition of energy describes correctly the chemical desorption process on bare surfaces. On icy surfaces, the chemical desorption process is much less efficient and a better description of the interaction with the surface is still needed.}{We show that the mechanism that directly transforms solid species to gas phase species is efficient for many reactions.}


\titlerunning{Dust as interstellar catalyst I}
\maketitle

\section{Introduction}
The chemical composition of regions where stars are forming influences the thermal balance and the evolution of interstellar clouds on their way to form stars (\citealt{hocuk2014}). Recent observations of cold environments (i.e. pre-stellar cores) have shown unexpected amount of some molecules in the gas phase (\citealt{Bacmann2012}), while these species should be depleted from the gas phase and adsorbed on cold dust grain (T$<$20 K) in physisorbed states (\citealt{collings2004}). 

In order to understand these observations, the transition between solid and gas phases has been studied in laboratory under physical-chemical conditions present in the ISM. In particular, many studies have been focussing on thermal desorption (\citealt{bisschop2006, noble15}) and non-thermal processes, like photo-desorption (\citealt{desimone2013}; \citealt{bertin2013}), sputtering (\citealt{Johnson2013}; \citealt{cassidy2013}), cosmic ray (CR) desorption (\citealt{leger1985}; \citealt{hasegawa1993}; \citealt{ivlev2015}) and chemical desorption (\citealt{cazaux2010}; \citealt{dulieu2013}; \citealt{minissale2014}).
  
Differently from the other processes, chemical desorption links solid and gas phases without the intervention of any external agents like photons, electrons or other energetic particles. In other words, it could be efficient in UV shielded and low CR environments where photo-desorption, CR (thermal) desorption or sputtering cannot be efficient.
The chemical desorption process starts from the energy excess of some radical-radical reactions. As described in \cite{minissaledulieu2014}, its efficiency depends essentially on four parameters: enthalpy of formation, degrees of freedom, binding energy and mass of newly formed molecules.

The aim of this article is to quantify the chemical desorption mechanism (part I) and its impact on interstellar gas (part II). 
The paper is organized as follow: In Sec. II, the experimental protocol and results are presented. In Sec.~III, we derive a law to quantify the chemical desorption mechanism. In Sec.~IV we discuss the main results of this work.


\section{Experiments}
The experiments have been performed using the FORMOLISM set-up (\citealt{amiaud2006}; \citealt{congiu2012}). In an ultra high vacuum chamber a gold mirror covered with amorphous silicates, oxidized graphite or compact amorphous water substrate is held at very low temperatures ($\sim$ 10~K). All the experiments are performed in sub-monolayer regime, to be sure that the interaction concerns bare surfaces, and is not perturbed by any other layer properties. The surface coverage is known using the specific desorption properties of the second layer (\citealt{noble2012}). A typical experiment proceeds as follow: one layer (or less) of non self-reactive molecules is deposited at low temperature. Just after its deposition another beam is aimed at the solid sample and the desorption flux is monitored simultaneously with a movable quadrupole mass spectrometer, placed 3 mm away from the surface. These experiments are refereed to as \rm{During Exposure Desorption}\rm\ (DED). 

To deal with radical-radical systems we sometimes use two simultaneous beams (i.e. [O+H] system ) because each separate beams will lead to a complete recombination (i.e. H+H$\longrightarrow$ H$_2$) and so a sequential deposition is meaningless.
The DED technique collect species in the gas phase during the reactive phase (beam exposure). Due to poor angular collection, signals are weak but are unambiguously due to the reactive systems. For example, if mass 20 is measured during the irradiation of an O$_2$ layer ( O$_2$ with mass of 32 a.m.u.) by D  atoms (mass 2), it is a direct evidence that D$_2$O is produced and undergoes chemical desorption. Such a signal is presented in the Figure 6 of \cite{chaabouni2012}. Mass confusions with cracking patterns of the QMS are still possible but can be estimated by careful calibration.
DED measurements allow to determine the coarse efficiency of the chemical desorption process for many considered reactions. Even though it is a direct measurement, most of the reactive systems are actually composed of many reactions. In the former example, it is clear that the O$_2$ + D reaction cannot directly produce D$_2$O, even if this signal appears as soon as the D exposure begins. For the case of water formation, more than 10 reactions are taking place simultaneously (see the Figure 7 of \cite{chaabouni12} for a graphical description). DED is a direct measurement of products, but the link between one product and one reaction is not necessary direct. 

The DED technique can be completed by indirect measurements of the chemical desorption. In a second phase, after the exposure, the surface temperature is increased and the desorption flux is monitored. This second phase, called temperature programmed desorption (TPD) experiments, allows us to determine the amount of formed products that did not desorb upon formation. TPD experiment is therefore an indirect process to measure the loss of surface species due to chemical desorption.

\subsection{How to determine the chemical desorption efficiency: the test case of the O+O system}
\begin{figure}
\centering
\includegraphics[width=8.8cm]{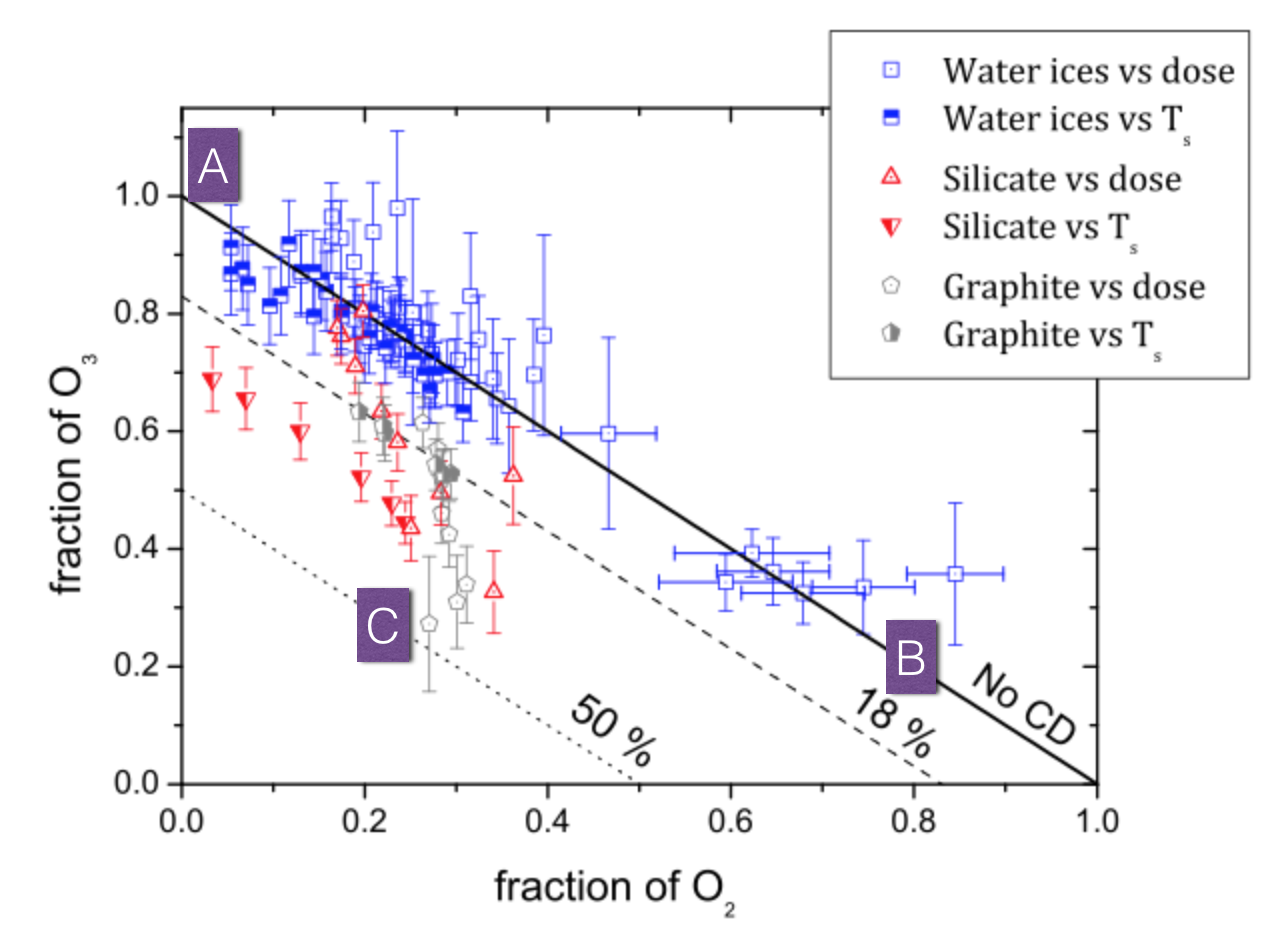}
\caption{Normalized fraction of O$_2$ and O$_3$ for the [O+O] reactive system obtained after TPD. The filled symbols represent the surface temperature variation ( between 6.5~K and 25~K ) while the empty symbols represent the dose variation (between 0.05~ML, and 1.0~ML). The black dots represent the data of \cite{minissaledulieu2014} where the method is detailed. Other points are coming from \cite{minissale2013}, \cite{minissale2014}, \cite{congiu2014} or are original data. The solid black line represent the zone of the plot where no chemical desorption is measured.}
\label{O2O3}
\end{figure}

In a recent experiment set of experiments \citep{minissale2014,congiu2014} we studied the formation of ozone and O$_2$ on different surfaces obtained by deposition of a unique beam composed of 85$\%$ of O and 15$\%$ of O$_2$. We choose this example because contrarily to most of the other reactive systems, there are here only 2 efficient reactions O+O $\longrightarrow$ O$_2$ and O+O$_2$ $\longrightarrow$ O$_3$. Therefore, there is no ambiguity or multiple possible assignment in the chemical reaction network. Each product corresponds to one reaction.
At the end of the exposure, the temperature of the surface has been increased and the amount of O$_2$ and O$_3$ (TPD) that were present on the surface could be measured. \figurename~\ref{O2O3} reports the results for experiments performed on water ice (squares), silicates (triangles) and graphitic (circles) surfaces. Keeping the same total exposure beam, the experiments can be repeated with various substrate temperature (so called ``vs T$_s$ experiments''), or keeping the same surface temperature (usually 10K), the exposure time can be varied, so the effect of the deposited dose can be studied (so called``vs dose''). In all the experiments, all the atoms are consumed during the exposure phase and we collect the O$_2$ and O$_3$ signals. The numbers reported in the X and Y axes are the fraction of O$_2$ and O$_3$ measured in the TPD. The point A on the  \figurename~\ref{O2O3} (coordinate 0, 1) represents an experiment where almost all the O/ O$_2$ deposited has been transformed into Ozone. This happens in high surface temperature conditions where the mobility of O atoms is high and thus any atoms has the possibility to react with an O$_2$ molecule. The point B (0.8, 0.2), corresponds to the case where 80$\%$ of the products have desorbed under the form of O$_2$. This is the case of very low temperature and coverage. In these conditions, the O+O reaction dominate over the O$_2$+O reaction. If the total fraction of O$_2$ and O$_3$ is equal to 1 (the sum of coordinates), then all the oxygen species sent on the surface are desorbing from the substrate during the TPD. This case is represented by the thick line which crosses both O$_2$ and O$_3$ fraction origin axis at coordinate 1. On the contrary, the point C (0.25,0.25) on the figure represents a case where the amount of species desorbing from the surface after the deposition is about equal to half of the  total amount of species that were deposited. The oxygen species missing from the TPD experiments are attributed to the chemical desorption process occurring during the exposure phase. The general conclusion drawn from \figurename~\ref{O2O3} is that the chemical desorption process can be efficient under some specific conditions.

We notice that experiments made on water substrates (blue points) all lie around the thick line representing the absence of chemical desorption.
The  reactions occurring on water ice surfaces lead to products remaining on the surface, and the chemical desorption process is almost negligible, or inside the error bars. Experiments made on  silicates and graphite surfaces, however,  show a large deficit in oxygen species. The experimental results show that almost 50$\%$ of the oxygen is missing from the surface for reactions occurring on graphite surfaces, while this fraction is of 30$\%$ for reaction on silicates and is negligible on water ices. For graphite and silicates surfaces, the amount of oxygen missing can vary with the coverage of the species on the surface (open triangles for silicates and open circles for graphite), which indicates that the chemical desorption process strongly depends on the surface coverage (\citealt{minissaledulieu2014}). Actually, the presence on the surface of a neighbor of similar mass increases strongly the probability of energy transfer. The energy detained by the molecule upon its formation is quenched very efficiently. In this case, the chemical desorption process is minimized.

\begin{figure}
\centering
\includegraphics[width=8.8cm]{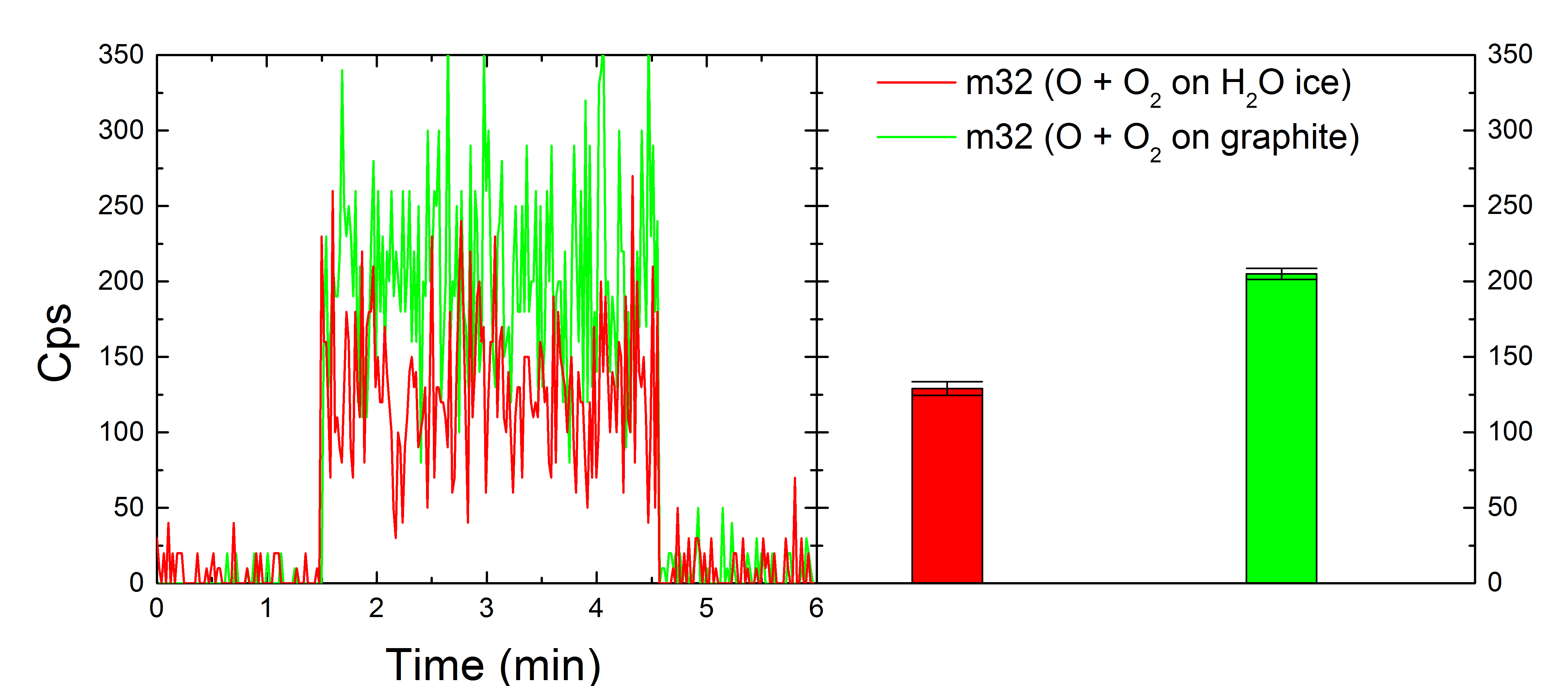}
\caption{Left panel: Raw DED measurements of mass 32  (O$_2$) during O/O$_2$ deposition (between 1.5 and 4.5 min) on graphite (green) and water ice (red) held at 10 K. Right panel: mean DED measurements over the 3 min exposure with associated error bars.
}
\label{DED}
\end{figure}

The oxygen species missing from the TPD experiments are attributed to the chemical desorption process. In order to confirm that the chemical desorption process is at the origin at the loss observed in the TPD, DED measurements are performed. This technique consists in measuring directly the increase of the partial pressure of the gas during the exposure of the reactants. For these type of measurements, the quadrupole mass spectrometer (QMS) head is close to the surface but still allows the deposition of species on the surface. This also implies that the position of the QMS is not optimized for the detection of the desorbing species. Furthermore, molecules directly formed may have a high kinetic energy, which reduces their interaction time passing through the QMS head. This mostly explains why the DED signals are low. On the left panel of the \figurename~\ref{DED}, we show the raw data measured during the exposure of O atoms on silicates and water ice surfaces. Before exposure (for time $\le$1.5 min), and after exposure (for time $\ge$4.5 min), the signal at mass 32 (a.m.u.) is very low. The signal during exposure, located between 1.5 and 4.5 min, is stronger for O$_2$ desorbing from a graphite surface than from a water ice surface. This is exactly the opposite situation to what we found in the TPD. This shows that the missing oxygen from our TPD are released into the gas during exposure, through the chemical desorption process, and that this process strongly depends on the surface on which reactions occur. The chemical desorption is found to be less efficient on water substrate than on oxidized graphite, via direct measurement (DED) or indirect measurement (TPD). 
Note that for the specific case of DED of a beam of O, a part of the O$_2$ beam is not dissociated and is detected in the QMS, and therefore does not come from the formation and desorption of O$_2$ on the surface. The DED technique provides good qualitative information, but is not suitable for quantitative analyses. It is however possible for us to sort qualitatively desorbing species. The species that we easily detect by DED should have an efficiency of few tens of \%, whereas those close to our detection limit, should be in the 5\%- 15\% range, depending on the level of background noise.

By combining TPD and DED experiments for O$_2$ on several surfaces, we could obtain an accurate measurement of the chemical desorption efficiency for the [O+O] system. This is possible because of the limited number of reactions occurring in this system. We studied many other systems. Some are quite simple like [N+N] (or [H+H]), where it is reasonable to think that there is only one reaction N+N $\longrightarrow$ N$_2$. But most other systems are more complex.

\begin{table*}[t]
\centering
\caption{List of experimental chemical desorption efficiencies for different reactions on three different surfaces (np-ASW, amorphous silicate, and oxidized graphite). For the cases in which we were not able to measure quantitatively the CD efficiency, we provide upper or lower limits. We list theoretical CD efficiencies calculated for an oxidized graphite.}\label{table1}
\resizebox{\textwidth}{!}{%
\begin{tabular}{c c c||cc||cc||c||ccc||c}
\rowcolor[gray]{0.85}
Experiments & Surface & Coverage & \multicolumn{2}{>{\columncolor[gray]{.85}}c}{Method} & Reaction&Desorbing&Delta H$_R$&\multicolumn{3}{>{\columncolor[gray]{.85}}c}{Experimental CD efficiency} &Theoretical CD efficiency\\
\rowcolor[gray]{0.85}
& Temperature & Range & DED & TPD & & product & &  & &  &  \\
\rowcolor[gray]{0.9}
& & & & & &  &  &np-ASW & Amorphous  & Oxidized  &  \\
\rowcolor[gray]{0.9}
& & &  &  &  &  & & &Silicate &  HOPG & \\
\rowcolor[gray]{0.85}
& (K) & (ML) & & &   & &eV&\%&\%&\%&\% \\
 \rowcolor[gray]{0.96}
   &     &              &                                    &   & O+H      & OH  &4.44&   25$\pm$15{\tiny$^{*}$}    & -- &  50$\pm$25{\tiny$^{*}$} & 39\\
\rowcolor[gray]{0.96}
\multirow{-2}{*}{[O+H]} & \multirow{-2}{*}{10} & \multirow{-2}{*}{$<$1} & \multirow{-2}{*}{\textit{$\surd$}} & \multirow{-2}{*}{\textit{$\surd$}}    & OH+H     & H$_2$O      & 5.17&   30$\pm$15{\tiny$^{*}$}    & -- & 50$\pm$25{\tiny$^{*}$}  & 27\\
\rowcolor[gray]{0.94}
     &  & &           &         & OH+H     & H$_2$O  &5.17&   $<$40$\pm$20{\tiny$^{a}$}     & $<$70$\pm$20{\tiny$^{a}$}  &  $<$80$\pm$20{\tiny$^{b}$} &27 \\
\rowcolor[gray]{0.94}
    &    & &             &      & O$_2$+H  & O$_2$H      &   2.24& $<$8{\tiny$^{*}$}   & 10$\pm$10{\tiny$^{b}$}& -- &1.4\\
\rowcolor[gray]{0.94}
  &      & &           &        & O$_2$H+H & H$_2$O$_2$  & 3.69&   $<$8{\tiny$^{*}$}   & $<$5{\tiny$^{b}$}     &  -- &0.5\\
\rowcolor[gray]{0.94}
     &    & &          &        & O$_2$H+H & 2 OH        &  1.47& $<$8{\tiny$^{*}$}    &$<$5{\tiny$^{b}$}     &  --&0.3\\
\rowcolor[gray]{0.94}
\multirow{-5}{*}{[O$_2$+H]} & \multirow{-5}{*}{10} & \multirow{-5}{*}{0.5-1}  & \multirow{-5}{*}{\textit{$\surd$}} &\multirow{-5}{*}{\textit{$\surd$}}     & H$_2$O$_2$+H& H$_2$O+OH& 2.95& $<$5{\tiny$^{*}$}    & $<$5{\tiny$^{b}$} & --&2.1\\
\rowcolor[gray]{0.96}
   & &  &       &       & O$_3$+H    & O$_2$+OH  &3.33&         -- &  $<$10{\tiny$^{b}$} & $<$8{\tiny$^{*}$} &8 \\
\rowcolor[gray]{0.96}
\multirow{-2}{*}{[O$_3$+H]} & \multirow{-2}{*}{10 and 45} & \multirow{-2}{*}{0.5-1}& \multirow{-2}{*}{\textit{$\surd$}} &\multirow{-2}{*}{\textit{$\surd$}}              & OH+H    & H$_2$O  &  5.17&    --   &  80$\pm$20{\tiny$^{b}$} & -- &27\\
\rowcolor[gray]{0.94}
     &   & &     &              & O+O        & O$_2$     & 5.16&$<$5{\tiny$^{c}$}   & 40$\pm$10$^{\#}${\tiny$^{d}$}   & 80$\pm$10$^{\#}${\tiny$^{e}$}&68\\
\rowcolor[gray]{0.94}
\multirow{-2}{*}{[O+O]} & \multirow{-2}{*}{8 to 25} &\multirow{-2}{*}{$<$1}& \multirow{-2}{*}{\textit{$\surd$}} & \multirow{-2}{*}{\textit{$\surd$}}    & O$_2$+O    & O$_3$     & 1.1&$<$5{\tiny$^{c}$}   & $<$5{\tiny$^{d}$} & $<$5{\tiny$^{e}$}& 0\\
\rowcolor[gray]{0.96}
\multirow{-2}{*}{[N+N]} &  \multirow{-2}{*}{10} & \multirow{-2}{*}{$<$1}& \textit{$\surd$} &  \textit{$\surd$}               & N+N    & N$_2$  & 9.79&$>$50$^{\&}${\tiny$^{*}$}   & $>$70$^{\&}${\tiny$^{*}$} & $>$70$^{\&}${\tiny$^{*}$}& 89\\
\rowcolor[gray]{0.94}
        & &  &       &        & CO+H & HCO &0.66& --   & -- & 10$\pm$8{\tiny$^{*}$}& 0.7\\
\rowcolor[gray]{0.94}
      & & &          &        & HCO+H& CO+H$_2$  &3.85& --  & -- & 40$\pm$20{\tiny$^{*}$} & 47\\
\rowcolor[gray]{0.94}
\multirow{-3}{*}{[CO+H]} & \multirow{-3}{*}{10}& \multirow{-3}{*}{0.8-2.5} & \multirow{-3}{*}{\textit{$\surd$}} & \multirow{-3}{*}{\textit{$\surd$}} & HCO+H& H$_2$CO   & 3.91&--   & -- & $<$8{\tiny$^{*}$} &7\\
\rowcolor[gray]{0.96}
       &   & &    &        & H$_2$CO+H & CH$_3$O  & 0.88& --  & -- & $<$8{\tiny$^{*}$}&0\\
\rowcolor[gray]{0.96}
        &  & &     &        & H$_2$CO+H& HCO+H$_2$  & 0.61& -- & -- & 10$\pm$5{\tiny$^{*}$}& 0\\
\rowcolor[gray]{0.96}
        &  & &     &         & HCO+H& CO+H$_2$&3.85& --  & -- & 40$\pm$20{\tiny$^{*}$} &47\\
\rowcolor[gray]{0.96}
        &  & &     &          & HCO+H& H$_2$CO &3.91&  --  & -- & 10$\pm$5{\tiny$^{*}$} &7\\ 
\rowcolor[gray]{0.96}
\multirow{-5}{*}{[H$_2$CO+H]}&\multirow{-5}{*}{10 and 55} & \multirow{-5}{*}{0.8-2} & \multirow{-5}{*}{\textit{$\surd$}}  &\multirow{-5}{*}{\textit{$\surd$}}  & CH$_3$O+H& CH$_3$OH   &4.56& $<$8{\tiny$^{*}$}   & --& $<$8{\tiny$^{*}$} &2.3\\
\rowcolor[gray]{0.94}
\multirow{-1}{*}{[Ar+H]}   & \multirow{-1}{*}{10}   &  \multirow{-1}{*}{1}  & \textit{$\surd$}  &\textit{$\surd$}         & Ar+H & Ar &&  --  & $<$5{\tiny$^{*}$} & -- &\\
\rowcolor[gray]{0.96}
\multirow{-1}{*}{[NO+H/O/N]}   &   \multirow{-1}{*}{10}  & \multirow{-1}{*}{0.5-5}   & \textit{X}  &\textit{$\surd$}     &   many   & many && $<$8{\tiny$^{*}$}   & $<$8{\tiny$^{*}$} & $<$8{\tiny$^{*}$} &\\
\rowcolor[gray]{0.94}
\multirow{-1}{*}{[CO+O]} & \multirow{-1}{*}{10}  & \multirow{-1}{*}{0.5-4}   & \textit{$\surd$}  &\textit{$\surd$}                     & CO+O& CO$_2$ & 5.51& $<$5{\tiny$^{*}$}   & -- & $<$5{\tiny$^{*}$}&22\\
\rowcolor[gray]{0.96}
\multirow{-1}{*}{[H$_2$CO+O]} & \multirow{-1}{*}{10 and 55}   & \multirow{-1}{*}{0.8-2}  & \textit{X}  &\textit{$\surd$}                     & H$_2$CO+O& CO$_2$+H$_2$ &5.45& $<$10{\tiny$^{*}$}   & -- & $<$10{\tiny$^{*}$}&8\\
\rowcolor[gray]{0.94}
\multirow{-1}{*}{[CH$_3$OH+H]} &\multirow{-1}{*}{10}  & \multirow{-1}{*}{0.5-2}  & \textit{$\surd$}  &\textit{$\surd$}      & CH$_3$OH+H& CH$_3$OH & &$<$8{\tiny$^{*}$}   & -- & $<$8{\tiny$^{*}$}& \\
\rowcolor[gray]{0.96}
\multirow{-1}{*}{[CH$_3$OH+O]}  & \multirow{-1}{*}{10}  & \multirow{-1}{*}{0.5-2} & \textit{$\surd$}  &\textit{$\surd$}      & CH$_3$OH+O& CH$_3$OH & &$<$8{\tiny$^{*}$}   & -- & $<$8{\tiny$^{*}$}&\\
\end{tabular}}

{\textit{Surface temperature} indicates deposition temperature of both reactants. \textit{Coverage range} indicates the coverage of molecular ice growth before atoms irradiation; in the case of experiments performed only with atomic species, \textit{coverage range} indicates the total amount of species sent on the surface.}
\textit{$\surd$} and \textit{X} are used to indicate if the experimental procedure (DED or TPD) have been used or not, respectively; 
$^{\#}$ CD decreases as a function of coverage;
$^{\&}$ experiments performed with excited particles (see ~\citealt{minissalethesis2014});
$^{a}$ \citealt{chaabouni2012};
$^{b}$ \citealt{dulieu2013}; 
$^{c}$ \citealt{minissale2013}; 
$^{d}$ \citealt{minissale2014}; 
$^{e}$ \citealt{minissaledulieu2014};
$^{*}$ This work.  
\end{table*}
\subsection{How to determine the chemical desorption efficiency: more complex systems}
Table~\ref{table1} lists the different experiments performed as well as the main reactions involved in each experiment. In the case of the [O+H] experiment, there are 2 main reactions occurring in the experiment, O+H $\longrightarrow$ OH and OH+H $\longrightarrow$ H$_2$O that we can measure with TPD and DED. To derive the chemical desorption efficiency for the reaction OH+H $\longrightarrow$ H$_2$O, we assume that the fraction of H$_2$O detected is due to the OH molecules that stayed on the surface upon formation. The amount of chemical desorbed H$_2$O observed in this experiment indirectly gives information on the amount of OH staying on the surface or desorbed upon formation. We positively detected OH for the [O+H] system, but not for the [O$_2$+H] nor [O$_3$+H] systems. By looking at the possible reaction network of water, we can attribute the high chemical desorption efficiency of OH to the O+H reaction. It is not trivial since other routes can produce OH, such as H$_2$O$_2$+H$\longrightarrow$ H$_2$O+OH or O$_2$H+H$\longrightarrow$ 2 OH.

To summarize, by combining TPD and DED experiments, one can identify which are the ``missing species'' by using a small reaction network. This allows us to derive some estimates of the chemical desorption efficiency for the formation of different species on surfaces. Here we present several experiments which are listed in Table~\ref{table1}. For each experiment, as mentioned previously, more than one reaction can occur leading to different products with distinct chemical desorption efficiencies. These reactions, as well as the enthalpy of reaction, the chemical desorption efficiency on water ice, amorphous silicates and oxidized graphite are reported in Table~\ref{table1}.

\section{Theoretical estimate of the desorption efficiency}
We propose to use a simple assumption of equipartition of energy to reproduce the chemical desorption observed experimentally. We suppose that the total energy budget $\Delta H_R$ (see Table\ref{table1}) is shared between all the degrees of freedom $N$=3$\times$ n$_{atoms}$. We use the generic 3 degree of freedom per atom. For a diatomic molecule, it would be decomposed in 3 degrees for translation, 1 degree for vibration, and 2 for rotation axis. Of course, the symmetry can affect this number slightly, however this would be a small correction. Among these degrees of freedom, only the kinetic energy perpendicular to the surface will allow the newly formed species to desorb, with the condition that this kinetic energy is more important than the binding energy ($E_{binding}$). The binding energies of the species considered in our network are reported in Table~\ref{bind}. Initially, the translational momentum is negligible since we believe that in most of our experiments the species admitted on the surface thermalize rapidly and the diffusion mechanism is the dominant process. To desorb from the surface, the newly formed products have to bounce against the surface to gain velocity in the direction perpendicular to the surface. The simplest way to treat this process is to assume an elastic collision,
\begin{equation}
\epsilon=\frac{(M-m)^2}{(M+m)^2}.
\end{equation}
Here $\epsilon$ is the fraction of kinetic energy retained by the product $m$ colliding with the surface which has an effective mass $M$. For the specific case of hyper thermal scattering of O$_2$ on graphite surface, \cite{hayes2012} have proposed an equivalent mass of 130 a.m.u. The effective mass is much higher than one single C atom, or even a carbon ring, because the rigidity of the surface induces a collective behavior of the atom and a larger effective mass. 

\begin{table}[h]
\caption{Binding energies on bare and icy surfaces. \label{bind}}
\begin{tabular}{lll}
Species & E$_{bare}$ &E$_{ice}$\\
\hline\hline
H&550$^{a,b}$&650$^c$\\
H$_2$&300$^{a,d}$&500$^e$\\
O&1500$^b$&1400$^f$\\
OH&4600$^g$&4600$^g$\\
H$_2$O&4800$^g$&4800$^h$\\
O$_2$&1250$^{i}$&1200$^i$\\
O$_3$&2100$^j$&2100$^j$\\
HO$_2$&4000$^g$&4000$^g$\\
H$_2$O$_2$&6000$^g$&6000$^g$\\
CO&1100$^i$&1300$^{i,k}$\\
CO$_2$&2300$^i$&2300$^{i,k}$\\
HCO&1600$^l$&1600$^l$\\
H$_2$CO&3700$^m$&3200$^{m}$\\
CH$_3$O&3700$^l$&3700$^l$\\
CH$_3$OH&3700$^n$&3700$^{n}$\\
N&720&720$^{f}$\\
N$_2$&790$^o$&1140$^p$\\
\hline\hline
\end{tabular}
\\
\rm{For HCO, CH$_3$O and CH$3$OH we assume  E$_{ice}$=E$_{bare}$.}\rm\ $^a$\citealt{cazaux2004}, $^b$\citealt{bergeron2008}, $^c$\citealt{Al-Halabi2007}, \citealt{amiaud2007}, $^d$\citealt{Pirronello1997}, $^e$\citealt{amiaud2006} for low coverage $^f$~Minissale et al. submitted, $^g$\citealt{dulieu2013}, $^h$\citealt{speedy1996}, \citealt{fraser2001} energy derived of 5800~K with pre-factor of 10$^{15}$ s$^{-1}$ here corrected to have pre-factor of 10$^{12}$ s$^{-1}$, $^i$\citealt{noble2012b}, $^j$\citealt{borget2001}, \citealt{minissalethesis2014},  $^k$\citealt{Karssemeijer2014},$^l$\citealt{garrod2006}, $^m$\citealt{noble2012}, $^n$\citealt{collings2004}, $^o$\citealt{fuchs2006}, $^p$~extrapolated from \citealt{kimmel2001}. 
\end{table}

\rm{The chemical energy is certainly splitted in many different ro-vibronic states, not at all in thermal equilibrium. We assume that an equal share of energy is spread over all the degrees of freedom, so that  the total chemical energy available for the kinetic energy perpendicular to the surface is $\epsilon \Delta H_R/N$. We assume that the only velocity perpendicular to the surface corresponds to a distribution of speed/temperature. In this sense, kT = $\epsilon \Delta H_R/N$.}\rm\ The probability for the product to have an energy (or liberation velocity) larger than the binding energy becomes:

\begin{equation}
CD = e^{-\frac{E_{binding}}{\epsilon \Delta H_R /N}}.
\end{equation}

\begin{figure}
\centering
\includegraphics[width=8.8cm]{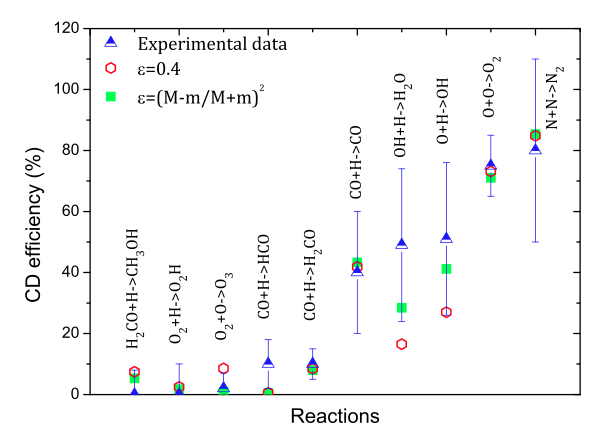}
\caption{Measured CD efficiencies in blue with error bars. In green, the result of the formulae (eq. 2) which include division among the degree of freedom and fraction of kinetic energy after bounce ($\epsilon$ is mass dependent). In red, equal share of energy and constant value of  $\epsilon$.}
\label{FigCD}
\end{figure}

In figure ~\ref{FigCD}, we compare the results obtained experimentally (blue triangles) to our predictions. We use first a constant $\epsilon$ of 0.4, which implies that we assume that 40$\%$ of the total energy is retained by the product (red hexagons). In this case, our predictions do not reproduce well the experimental results. In a second step, we consider a fraction $\epsilon$ which depends on the mass of the products, as described in eq. 1. These predictions are represented as green squares and do match very well the chemical desorption efficiencies observed experimentally. This clearly indicates that the chemical desorption process is sensitive to the binding energy, the energy distributed in the products (depending on their mass), and the degrees of freedom. For some of the reactions such as OH+H, O+H, our predictions slightly underestimate the experimental results. 

The formula (eq. 2) can be translated to the case of silicates surfaces. The efficiency is slightly smaller, and thus reducing the effective mass from 130 to 110 gives a very good match for the most constrained reaction (O+O). Taking into account the degree of inaccuracy of our experimental determination of CD, adopting a common value of 120 amu is appropriate to treat the trend of CD for bare grains.

The case of water substrate (np-ASW), which is even more important from the perspective of explaining the return in the gas phase of complex organic molecules (COMs), appears to be less simple. We have two contradictory results. The chemical desorption of O+O is at least 8 times less, but the OH+H reaction seems to be still efficient, even if lowered by a factor of $\sim$ 2. The mass argument does no longer stand. In the part II of our work, we assume that the chemical desorption efficiency on ices is our CD coefficient divided by 10 for the reactions that have not been constrained (CD$_{ice}$=CD$_{bare}$/10 while CD$_{ice}$(OH+H) = 25\%, CD$_{ice}$(O+H) = 30\% and CD$_{ice}$(N+N) = 50\%).

\section{Discussion}
Previously \cite{garrod2007}  have proposed in their models a formula to describe the chemical desorption (or reactive desorption). They derive the probability to undergo desorption in two steps. The first one is based upon RRKM estimation and takes into account energetic aspects (degree of freedom, enthalpy of reaction and binding energy). In most of the cases this coefficient is found to be high. The second step, interaction with the surface, is parametrized with an analytic formula. We first tried to use this approach, but unsuccessfully. The main problem, already mentioned by the authors, is that the first step (energetic aspects) is not really discriminating in case of exothermic reactions, whereas the second one, which is unknown, is the most important. The net result is that when this formula is used, there is no really discrimination among the reactions. However, we clearly demonstrated that O+O has a high probability to desorb and O$_2$+O has not, while both of these reactions are exothermic. This a strong argument to reject the former approach.

\begin{figure}
\centering
\includegraphics[width=8.8cm]{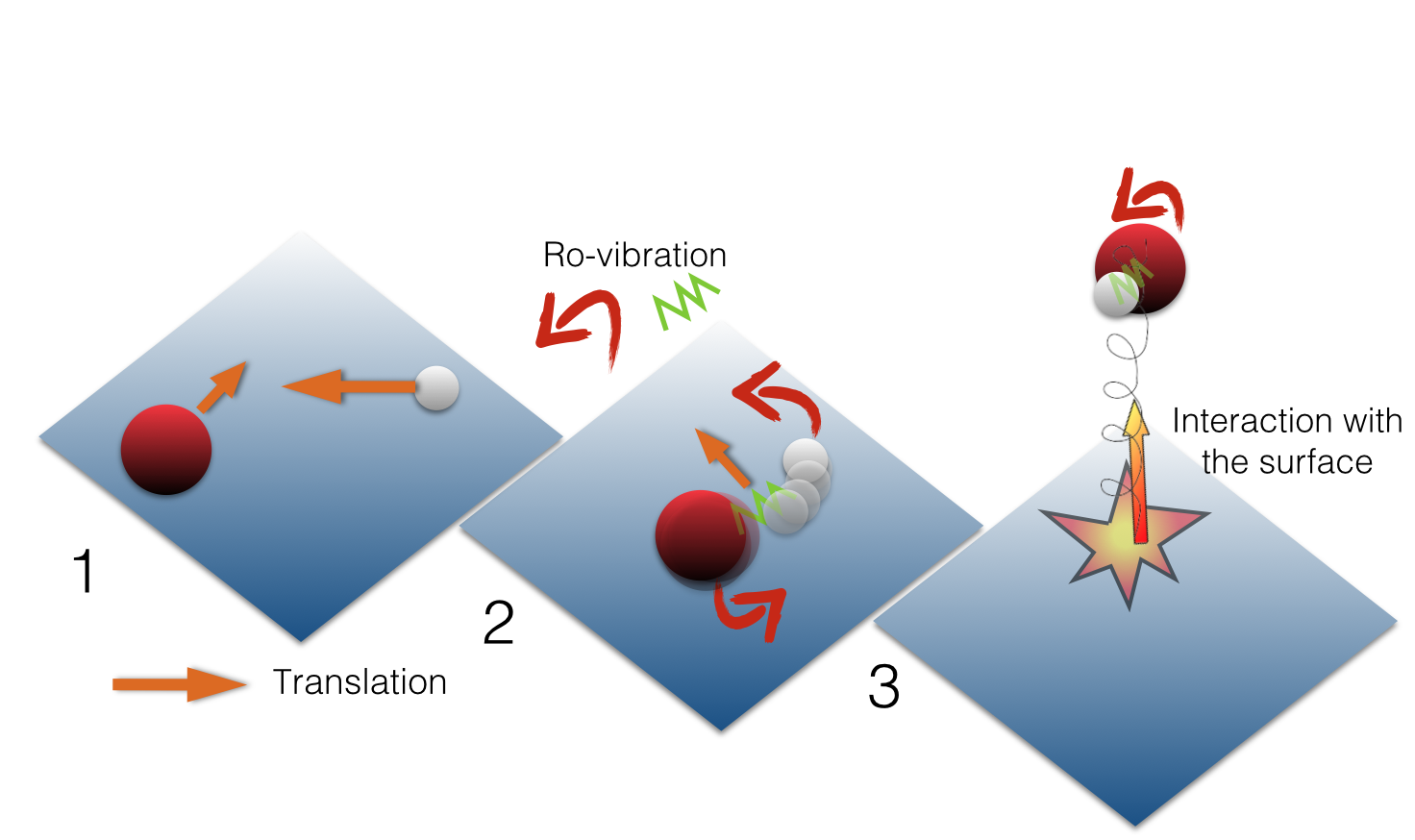}
\caption{Illustration of the chemical desorption process. After an initial meeting (1), the molecule is highly excited (2) and converts a fraction of its energy into vertical motion, thanks to the interaction with the surface (3)}
\label{Figbounce}
\end{figure}

Our simple hypothesis of equipartition of energy, coupled with elastic collision with the surface, reproduces the observations well. Although it is an empirical approach, it allows to sort among the products that should undergo the chemical desorption. The basic idea underlying our proposition is inspired by dynamic calculations in specific systems such as H+H on graphite \citep{morisset2004, bachellerie2007} or O+H on graphite \citep{bergeron2008}. We have illustrated the mechanism of this scenario in \figurename~\ref{Figbounce}. The first step consists in a collinear approach parallel to the surface. When the molecule is formed, the trajectories of both atoms adopt a more disordered motion symptomatic of the high internal energy in vibration and rotation of the newly formed molecules, up to a point where the molecule interacts with the surface and that a part of the motion is transferred perpendicularly to the surface provoking the desorption. 

We point out that the chemical desorption process seems to favor small molecules with low binding energies. Therefore, this process may be part of the reason COMS are present in dark clouds. We notice that our prediction of around one percent of chemical desorption for methanol would be sufficient to sustain the observed value, as modeled by \cite{vasyunin2013}. The second part of our paper will examine the impact of the derivation of our formulae in an astrophysical context.

The reduction of the chemical desorption efficiency on water ice substrates is somehow to be expected. Similar behavior has been observed in photodesorption of CO. Pure CO ices exhibit an efficient wavelength dependent photodesorption (up to one percent per incident photon, \citealt{bertin2013}) or may be a less by a factor 3-11 \citep{hemert2015}. On water substrate some authors have estimated the CO photodesorption to be around 10$^{-3}$ \citep{oberg2009} or one order of magnitude less (\citealt{desimone2013}). \rm{However, even if the CO photodesorption process is not very well constrained, it is clear that the water substrate prevents or dramatically reduces its efficiency.}\rm\ This probably is due to the very fast transportation of energy in the water network and to the reduction of the indirect photodesorption. 

In the case of the chemical desorption on water substrate, the weakness of our coarse model lies probably in the assumption of elastic collisions, which are not adapted to the case of water. Many more experimental investigations, as well as detailed computations would help to solve this problem. Finally, we note that even if water ice substrate seems to be a good trapping environment, and can dramatically reduce the chemical desorption efficiency, in the central part of dark clouds, CO layers are expected to cover an inner water layer around the solid seeds of dust. Therefore, another set of experiments should be performed using a CO ice substrate.

\section{Conclusions}
In this study we present a collection of experiments in order to quantify which fraction of species forming on surfaces is ejected into the gas phase upon formation. This process, called chemical desorption, has been observed experimentally (\citealt{dulieu2013}). With this work we derive an analytical expression to quantify the chemical desorption process on different substrates such as oxidized graphite and amorphous silicate. This formula, which depends on the equipartition of the energy of newly formed products and the binding energy of the products, reproduces well the experimental results on bare surfaces. However, on icy surfaces the chemical desorption process is strongly reduced and cannot be well quantified by our experimental results. 

In part II of our study, we address the importance of the chemical desorption process on the composition of interstellar gas, by combining gas phase chemistry as well as surface and bulk chemistry, and using our analytical formula to account for the chemical desorption process.

\begin{acknowledgements}
We acknowledge the support of the French National PCMI program funded by the CNRS and the support of the DIM ACAV, a funding program of the R\'egion Ile de France. S. C. is supported by the Netherlands Organization for Scientific Research (NWO; VIDI project 639.042.017) and by the European Research Concil (ERC; project PALs 320620). We would like to thank the anonymous referee for his/her comments and suggestions that greatly improved our manuscript.
\end{acknowledgements}

\end{document}